\documentstyle[prl,aps,floats,twocolumn,epsf]{revtex}

\begin{document}

\par
\begingroup
\twocolumn[%
{\large\bf\centering\ignorespaces
Detectability of inflation-produced gravitational waves \vskip2.5pt}
{\dimen0=-\prevdepth \advance\dimen0 by23pt
\nointerlineskip \rm\centering
\vrule height\dimen0 width0pt\relax\ignorespaces
Michael S. Turner
\par}
{\small\it\centering\ignorespaces
Departments of Physics and Astronomy \& Astrophysics, Enrico
Fermi Institute, University of Chicago, Chicago, IL 60637-1433\\
NASA/Fermilab Astrophysics Center,
Fermi National Accelerator Laboratory, Batavia, IL 60615-0500\\ \par}
{\small\rm\centering(\ignorespaces 12 July 1996\unskip)\par}

\par
\bgroup
\leftskip=0.10753\textwidth \rightskip\leftskip
\dimen0=-\prevdepth \advance\dimen0 by17.5pt \nointerlineskip
\small\vrule width 0pt height\dimen0 \relax

Detection of the gravitational waves excited during inflation
as quantum mechanical fluctuations is a key test of inflation and
crucial to learning about the specifics of the inflationary model.
We discuss the potential of Cosmic Background Radiation (CBR)
anisotropy and polarization and of laser interferometers such
as LIGO, VIRGO/GEO and LISA to detect these gravity waves.

\par\egroup
\vskip2pc]
\thispagestyle{plain}
\endgroup

{\it Introduction}
Inflation addresses most of the fundamental problems in cosmology --
the origin of the flatness, large-scale smoothness, and
small density inhomogeneities needed to seed all the structure
seen in the Universe today.  If correct, it would
extend our understanding of the Universe to as early
as $10^{-32}\sec$ and open a window on physics at energies
of order $10^{15}\,$GeV.  However, at the moment there is little evidence
to confirm or to contradict inflation and no standard model of inflation.

The key to testing inflation is to focus on its three basic
predictions \cite{inflate}:
spatially flat Universe (total energy density equal to the
critical energy density); almost scale-invariant spectrum
of gaussian density perturbations \cite{scalar}; and
almost scale-invariant
spectrum of stochastic gravitational waves \cite{tensor}.
The first two predictions have important
implications:  the existence of nonbaryonic dark matter, as
big-bang nucleosynthesis precludes baryons from contribution
more than about 10\% of the critical density \cite{bbn}, and the cold dark
matter scenario for structure formation, based upon the idea
that the nonbaryonic dark matter is slowly moving elementary particles
left over from the earliest moments \cite{cdm,cdm2}.  A host
of cosmological observations are now beginning to sharply
test the first two predictions \cite{cdm2}.

Gravity waves are a telling test and probe of inflation:
They provide a consistency check (see below); they are essential to learning
about the scalar potential that drives inflation \cite{reconstruct};
and they are a compelling signature of
inflation -- both a flat Universe and scale-invariant density
perturbations were advocated before inflation.

Detecting inflation-produced gravity waves presents a great
experimental challenge \cite{lyth}.  In this {\it Letter} we
discuss the potential of CBR anisotropy or polarization
and of direct detection by the laser-interferometers to
test this key prediction of inflation.

{\it Quantum Fluctuations}  The (Fourier) spectra of metric fluctuations
excited during inflation are characterized by power laws in
wavenumber $k$, $k^{n}$ for density perturbations (scalar metric
fluctuations) and $k^{n_T-3}$
for gravity waves (tensor metric fluctuations).  Scale
invariance for density perturbations ($n=1$) corresponds to fluctuations
in the Newtonian potential that are independent of wavenumber; scale
invariance for gravity waves ($n_T = 0$) corresponds to dimensionless
horizon-crossing strain amplitudes that are independent of wavenumber.
The power-law indices are related to the scalar field
potential, $V(\phi )$, that drives inflation:
\begin{eqnarray}
n-1 & = & -{m_{\rm Pl}^2 \over 8\pi}\left( {V_*^\prime \over V_*}\right)^2
        +{m_{\rm Pl}^2\over 4\pi}\left( {V_*^\prime\over V_*}\right)^\prime ,\\
n_T & = & -{m_{\rm Pl}^2 \over 8\pi}\left( {V_*^\prime \over V_*}\right)^2 .
\end{eqnarray}
The overall amplitude of each spectrum can be characterized by
its contribution to the quadrupole anisotropy of the CBR,
\begin{eqnarray}
S  \equiv {5\langle |a^S_{2m}|^2\rangle \over 4\pi} & = &
        {2.2 (V_*/m_{\rm Pl}^4) \over (m_{\rm Pl} V_*^\prime /V_*)^2} , \\
T  \equiv {5\langle |a^T_{2m}|^2\rangle \over 4\pi} & = &
        0.61 (V_*/m_{\rm Pl}^4) ,
\end{eqnarray}
where $S$ refers to scalar and $T$ to tensor, $V_*$ is the value
of the inflationary potential when the present horizon scale ($k=H_0$)
crossed outside the Hubble radius during inflation, $V_*^\prime$ is
the first derivative of the potential at that point, and $m_{\rm Pl} =1.22\times
10^{19}\,$GeV \cite{formulae}.

There is a very important relation between
amplitude ($T/S$) and tilt ($n_T$), cf. Eqs.~(1-4),
\begin{equation}
n_T = -{1\over 7} \,{T\over S}.
\end{equation}
It not only provides a consistency check of inflation \cite{consis},
but it also has implications for the direct detection of gravity waves,
as it relates the overall amplitude to the tilt.  Note too, that
the tensor amplitude $T$ determines the value of the
inflationary potential, and together with $T/S$ and $n$,
the first two derivatives of the potential.  Any attempt
to reconstruct the inflationary potential requires knowledge
of the gravity-wave spectrum \cite{reconstruct}.

\begin{figure}[t]
\epsfxsize=3in
\centerline{\epsfbox{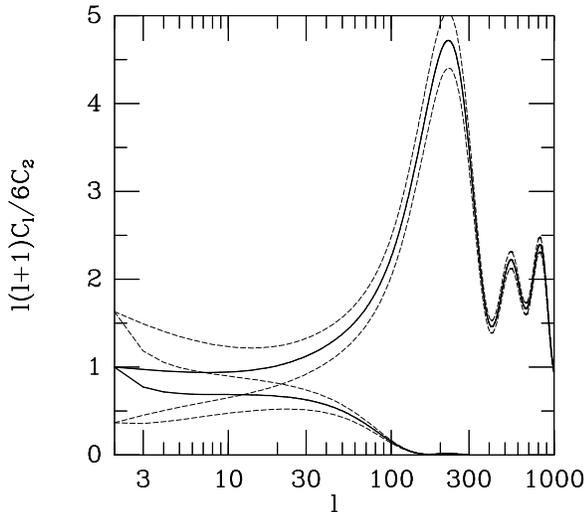}}
\caption{Angular power spectra ($C_l \equiv
\langle |a_{lm}|^2\rangle$) of
CBR anisotropy for gravity waves (lower curves) and density perturbations
(upper curves), normalized to the quadrupole anisotropy;
broken lines indicate sampling variance.
Temperature fluctuations measured on angular scale
$\theta$ are approximately, $(\delta T/T)_\theta \sim
\protect\sqrt{l(l+1)C_l/2\pi}$ 
with $l\sim 200^\circ /\theta$ (courtesy of M.~White and U.~Seljak).}
\end{figure}

{\it CBR}  Inflation-produced density
fluctuations and gravity waves each give rise to CBR anisotropy
and polarization, specified by their predictions for the variance of the
multipole amplitudes of anisotropy and polarization \cite{gaussian}.
The CBR signatures are very different:
the tensor angular power spectrum falls off quickly for $l > 100$
and its level of polarization is about 30 times greater for $l < 30$
(see Figs.~1 and 2).  However, there is a fundamental limit
to the accuracy with which the variance of the multipoles can be determined:
Because only $2l+1$ multiple amplitudes can be measured for a given $l$,
the variance can be estimated to a relative precision of
$1/\sqrt{l+{1/2}}$ (known as sampling, or cosmic, variance).

\begin{figure}[t]
\epsfxsize=3in
\centerline{\epsfbox{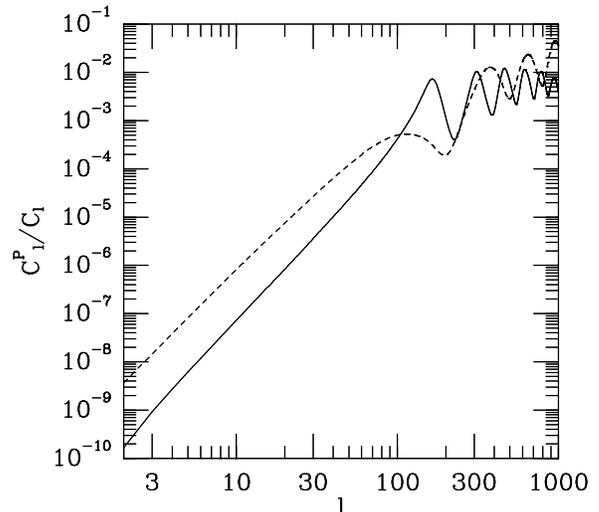}}
\caption{Polarization angular power spectra for gravity waves
(broken) and density perturbations (solid).
The polarization of the CBR anisotropy
is roughly $\protect\sqrt{C_l^P/C_l}$ (courtesy of M.~White and U.~Seljak).}
\end{figure}

Due to sampling variance $T/S$ must be greater than about
0.1 to ensure that the tensor signature of CBR anisotropy can be
detected \cite{knoxmst}.  In principle, polarization is
more promising -- $T/S$ as small as 0.02 could be detected
\cite{knoxmst}.  In practice, approaching this limit would
be extremely difficult,
requiring the polarization of the anisotropy to be measured with
0.01\% precision on large-angular scales.  Further, the polarization
on these scales is very sensitive to the ionization history of the Universe.

\begin{figure}[t]
\epsfxsize=3in
\centerline{\epsfbox{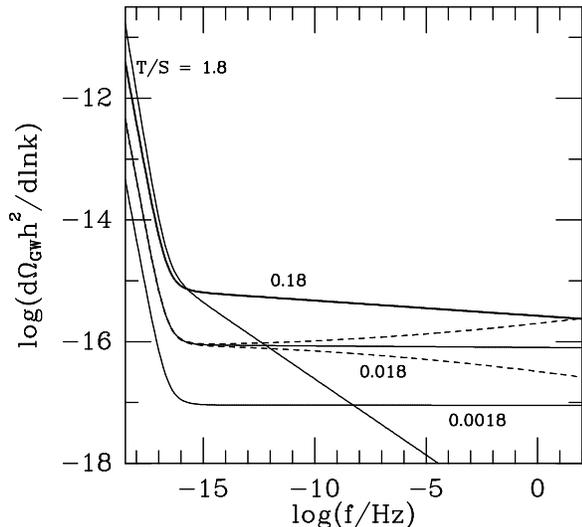}}
\caption{Spectral energy density in gravity
waves produced by inflation; for $T/S = 0.018$,
$dn_T/d\ln k = -10^{-3}$, 0, $10^{-3}$.  $T/S =0.18$ (heavy curve)
maximizes the energy density at $f=10^{-4}\,$Hz.
Curves are from Eq.~(\protect\ref{eq:energy})
using $H_0=60{\rm km\,s^{-1}\,Mpc^{-1}}$, $\Omega_0 =1$, and $g_*=3.36$.}
\end{figure}

{\it Direct Detection}  The inflation-produced background of
gravity waves offers at least one advantage -- the energy per
logarithmic frequency interval is roughly constant for
$f= 10^{-15}\,$Hz to $10^{15}\,$Hz (see Fig.~3),
\begin{equation}\label{eq:energy}
{d\Omega_{\rm GW}\over d\ln k} = {\Omega_0^2\,(V_*/ m_{\rm Pl}^4)\,\over
(k/H_0)^{2-n_T}} \left[ 1+{4\over 3}{k\over k_{\rm EQ}} +{5\over 2}
\left( {k\over k_{\rm EQ}} \right)^2 \right] ,
\end{equation}
where $k_{\rm EQ} = 6.22\times 10^{-2}\,{\rm Mpc^{-1}}\,
(\Omega_0h^2/\sqrt{g_*/3.36})$, is the scale that entered the horizon at
matter-radiation equality, $\Omega_0$ is the fraction of
critical density in matter (the balance of critical density is
assumed to be in vacuum energy), $\Omega_{\rm GW}$ is the
fraction of critical density in gravity waves,
wavenumber $k=2\pi f$, the Hubble constant $H_0 = 100h\,
{\rm km\,s^{-1}\,Mpc^{-1}}$, and $g_*$ counts the effective number of relativistic
degrees of freedom (3.36 for the CBR and three massless neutrino
species).  The factor in square brackets in Eq.~(\ref{eq:energy})
is a numerical fit to the transfer function for gravitational waves,
which accounts for the evolution of gravity-wave modes after they
re-enter the horizon (see Ref.~\cite{twl} for details).

The relationship between the tensor spectral index
and the overall amplitude can
be used to rewrite Eq.~(\ref{eq:energy})
in terms of $n_T$ (or $T/S$) alone.  Using
the fact that the variance of the CBR
quadrupole is given by the sum of the scalar and tensor
contributions ($Q=T+S$) and the COBE measurement,
$Q\simeq 4.4\times 10^{-11}$ \cite{cobe},
it follows that on the ``long plateau''
($k\gg k_{\rm EQ}$, $f\gg 10^{-15}\,$Hz)
\begin{eqnarray}\label{eq:sens}
{d\Omega_{\rm GW}h^2\over d\ln k} & = &
5.1\times 10^{-15} \,(g_*/3.36)\,{n_T\over n_T - 1/7} \nonumber \\
& \times & \exp [n_TN +{1\over 2} N^2 (dn_T/d\ln k) ],
\end{eqnarray}
where $N\equiv \ln (k/H_0)\simeq 33 +
\ln (f/10^{-4}{\rm Hz}) +\ln (0.6/h)$.
Note, if there are additional seas of relativistic
particles beyond the photons and three neutrino species
($g_*>3.36$), as has been advocated to improve the agreement
between the cold dark matter scenario and observations of
large-scale structure \cite{taucdm}, the energy density
in gravity waves is increased, perhaps by a factor of three \cite{easy}.

\begin{figure}[t]
\epsfxsize=3in
\centerline{\epsfbox{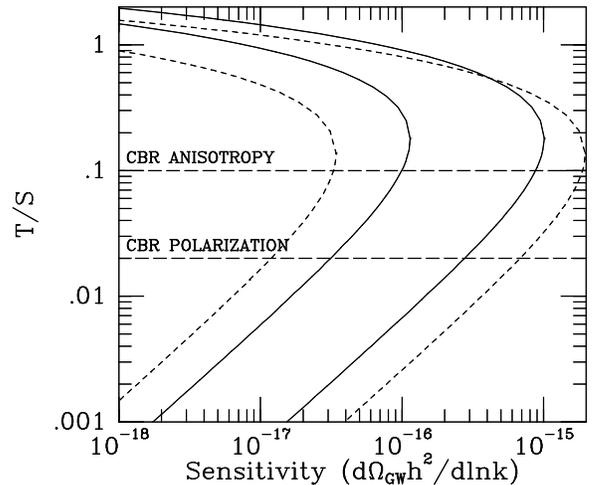}}
\caption{The range of $T/S$ probed
(interval interior to parabola) as a function of energy
sensitivity for $f=10^{-4}\,$Hz
(solid curves) and $f=100\,$Hz (broken curves).  The
``pessimistic'' (left) parabola assumes $dn_T/d\ln k = -10^{-3}$
and the ``optimistic'' (right) parabola assumes $dn_T/d\ln k =
10^{-3}$.  Also shown are the limiting sensitivity
of CBR anisotropy and polarization.}
\end{figure}

Since the spectrum is normalized at the Hubble scale ($k=H_0$)
and extrapolated to frequencies that
are some 15 orders of magnitude larger we have included the first
correction for the variation of the power-law index with scale.
The ``running'' of $n_T$ is given by \cite{komst},
\begin{equation}
{dn_T\over d\ln k} = -n_T[(n-1) - n_T] = -n_T\,{m_{\rm Pl}^2\over 4\pi}
\left( {V_*^\prime \over V_*} \right)^\prime .
\end{equation}
Typically, $dn_T/d\ln k \approx -10^{-3}$ \cite{komst};
it can be of either sign or even zero \cite{run}.  On a very optimistic
note, a CBR determination of $T/S$ and a laser interferometric
determination of the average spectral index (${\bar n_T} =
n_T+ 0.5 Ndn_T/d\ln k$) would allow the inference of $dn_T/d\ln k$.

An important
feature of Eq.~(\ref{eq:sens}) is the amplitude -- tilt relationship:
$n_T$ increases the prefactor, but tilts the spectrum
so as to decrease the amplitude at high frequencies.
At fixed frequency, the energy density
is maximized for $n_T = -\left( \sqrt{1+28/N} -1 \right)/14 \approx
-0.025$ ($f=10^{-4}\,$Hz).  Values for $n_T$ of this order are
realized in several models of inflation, e.g., chaotic inflation.

The energy density in a stochastic background of
gravitational waves can be expressed in terms of the {\it rms}
strain, $h_{rms}^2(k)\equiv k^3|h_{\bf k}|^2/2\pi^2$,
\begin{eqnarray}
{d\Omega_{\rm GW}\over d\ln k} & = &{2\pi^2\over3} \,\left({f\over H_0}
\right)^2\,h_{rms}^2(k) \nonumber \\
 &= &6.3h^{-2} \times 10^{-7}\,(f/{\rm Hz})^2
\,(h_{rms}/10^{-21})^2.
\end{eqnarray}
For fixed strain sensitivity, the energy-density sensitivity
varies with the square of the frequency because
$\rho_{\rm GW} \propto h_{rms}^2f^2$, and
so prospects for detection improve as $1/f^{2}$.

The range of $T/S$ accessible to a gravity-wave detector operating at
$f= 10^{-4}\,$Hz and $f= 100\,$Hz is shown as a function of energy
sensitivity in Fig.~4.  For either frequency, a sensitivity of
$d\Omega_{\rm GW}h^2/d\ln k \sim 10^{-15}$ is needed for a serious search for
inflation-produced gravity waves.  (The curves in Fig.~4 were computed
from Eq.~(\ref{eq:sens}) with $\Omega_0=1$ and $g_*=3.36$; for
$\Omega_0 <1$, only the labeling of the ordinate changes,
as the relation $T/S = -7n_T$, used to obtain the $T/S$ values,
is modified slightly \cite{mstwhite}.)

LIGO and the other detectors now being built
will operate at frequencies from 10\,Hz to several kHz, with initial
strain sensitivities of around $10^{-21}$, improving to $10^{-24}$
(at $f=10^2\,$Hz) \cite{ligo}.  Eq.~(\ref{eq:sens}) tells the sad story:
Even the most optimistic
estimate for LIGO's energy sensitivity misses the mark by
four orders of magnitude.   While Earth-based detectors cannot
operate at lower frequencies because of seismic noise, space-based
detectors can.   Early estimates indicated
that a strain sensitivity of slightly better than $h_{rms}
= 10^{-21}$ might be achieved at a frequency of $10^{-4}\,$Hz \cite{300yrs},
implying an energy sensitivity $d\Omega_{\rm GW}/d\ln k \sim 10^{-16}$,
sufficient to probe $T/S \sim 0.01$.  However, the design study for
LISA indicates an energy sensitivity of
around $d\Omega_{\rm GW}h^2 /d\ln k \sim 10^{-13}$, which misses
by two orders of magnitude \cite{lisa}.
(There is also a worrisome background of coalescing white-dwarf
binaries, which could dominate inflation at frequencies greater
than around $10^{-4}\,$Hz \cite{300yrs}.)

{\it Summary}  Gravity waves are an important prediction of inflation.
The CBR is sensitive to the longest-wavelength gravity waves ($10^{26}\,$cm to
$10^{28}\,$cm), but is fundamentally limited by sampling
variance.  The high-resolution ($l=2- 2000$) anisotropy maps
that will be made by two future satellite experiments,
MAP and COBRAS/SAMBA, might reach the sampling-variance limit,
$T/S\sim 0.1$.  Improving this by
polarization measurements does not look promising.
Laser interferometers are sensitive to much shorter wavelengths
($10^8\,$cm to $10^{13}\,$cm).  An energy sensitivity
$d\Omega_{\rm GW}h^2/d\ln k \sim 10^{-15}$
is required to search for the inflation-produced gravity-wave
background; a sensitivity of $10^{-16}$ opens the window
wide, perhaps allowing $T/S$ smaller than $0.01$ to be detected.
While Earth-based laser interferometers
are not likely to achieve this, there is some hope that
space-based detectors operating at low frequencies ($<10^{-4}\,$Hz) might.

We should temper our conclusions, which are based upon
the most accurate predictions available, with acknowledgment of
their limitations and our possible ignorance.  Assumptions
have been made:  one-field, slow-rollover inflation with a smooth
potential.  Nature could be more interesting.  If inflation
ends with the nucleation of bubbles there is an additional potent
source ($\Omega_{\rm GW}\sim 10^{-6}$) of gravitational waves
in a narrow frequency range \cite{vacuumpop};
pre-big-bang models predict a spectrum of gravity waves that rises
with frequency, making detection far more
promising \cite{prebb}; Grishchuk \cite{grishchuk} has long emphasized
the production of gravitational waves during the earliest moments
in a variety of scenarios.
Even if a sensitivity of $d\Omega_{\rm GW}h^2/
d\ln k \sim 10^{-15}$ cannot be achieved, it is still worth searching --
there could be surprises!

\bigskip\bigskip
We thank M. White for useful discussions.
This work was supported by the DoE (at Chicago and Fermilab) and
by the NASA (at Fermilab by grant NAG 5-2788).


\begin{references}
\bibitem{gaussian} Isotropy in the mean implies
that $\langle a_{lm} \rangle =0$; the gaussian nature of
the inflationary metric fluctuations implies that the multipole amplitudes
have gaussian distributions, fully specified by their predicted variances.
\bibitem{easy}  This is simple to understand:  the drop in
energy density from the Hubble scale to the plateau is given by the
redshift of matter -- radiation equality, which is inversely
proportional to the energy density in radiation.
\bibitem{run} Given the inflationary
potential the gravity-wave spectrum can be computed without assuming
a power law; we have done this for chaotic inflation models,
$V(\phi ) = a\phi^m$, to judge the accuracy of our approximation.
It is typically better than 33\%.  In addition,
since there is no standard model of inflation,
the use of $n_T$ and $dn_T/d\ln k$
to extrapolate the gravity-wave spectrum from very large scales
offers the advantage of generality.
\end{references}
\end{document}